\documentclass[]{pasj01_revised}
\usepackage{multicol}
\usepackage{multirow}

\begin{document} 

\title{Cosmology with the Square Kilometre Array by SKA-Japan}

\author{
Daisuke \textsc{Yamauchi}\altaffilmark{1,*}, 
Kiyotomo \textsc{Ichiki}\altaffilmark{2,3},
Kazunori \textsc{Kohri}\altaffilmark{4,5},
Toshiya \textsc{Namikawa}\altaffilmark{6,7},
Yoshihiko \textsc{Oyama}\altaffilmark{8},
Toyokazu \textsc{Sekiguchi}\altaffilmark{9},
Hayato \textsc{Shimabukuro}\altaffilmark{2,10},
Keitaro \textsc{Takahashi}\altaffilmark{10},
Tomo \textsc{Takahashi}\altaffilmark{11},
Shuichiro \textsc{Yokoyama}\altaffilmark{12},
Kohji \textsc{Yoshikawa}\altaffilmark{13}, 
on behalf of
SKA-Japan Consortium Cosmology Science Working Group
}

\altaffiltext{1}{Faculty of Engineering, Kanagawa University, Kanagawa, 221-8686, Japan}
\email{corresponding author; yamauchi@jindai.jp}
\altaffiltext{2}{Department of Physics and Astrophysics, Nagoya University, Aichi 464-8602, Japan}
\altaffiltext{3}{Kobayashi-Maskawa Institute for the Origin of Particles and the Universe, Nagoya University, Aichi 464-8602, Japan}
\altaffiltext{4}{Theory Center, IPNS, KEK, Tsukuba 305-0801, Japan}
\altaffiltext{5}{Sokendai, Tsukuba 305-0801, Japan}
\altaffiltext{6}{Department of Physics, Stanford University, Stanford, CA 94305, USA}
\altaffiltext{7}{Kavli Institute for Particle Astrophysics and Cosmology, SLAC National Accelerator Laboratory, Menlo Park, CA 94025, USA}
\altaffiltext{8}{Institute for Cosmic Ray Research, The University of Tokyo, Kashiwa, Chiba 277-8582, Japan}
\altaffiltext{9}{Institute for Basic Science, Center for Theoretical Physics of the Universe, Daejeon 34051, South Korea}
\altaffiltext{10}{Faculty of Science, Kumamoto University, 2-39-1 Kurokami, Kumamoto 860-8555, Japan}
\altaffiltext{11}{Department of Physics, Saga University, Saga 840-8502, Japan}
\altaffiltext{12}{Department of Physics, Rikkyo University, Tokyo 171-8501, Japan}
\altaffiltext{13}{Center for Computational Sciences, University of Tsukuba, Ibaraki 305–8577, Japan}



\maketitle

\begin{figure}
\vspace{-230mm}
\includegraphics[width=25mm]{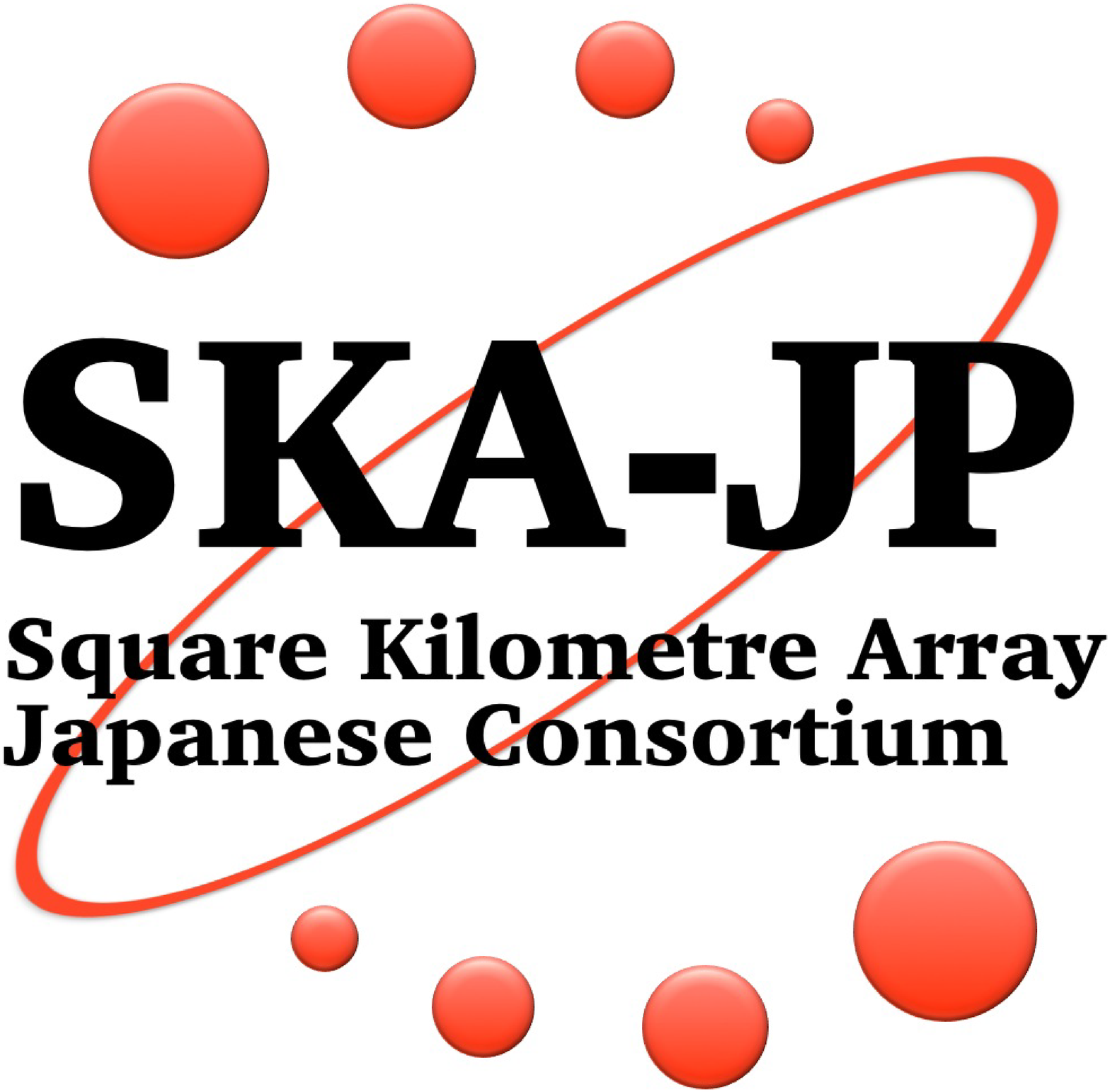}
\hspace{3.8cm}
\rightline{
RESCEU-2/16, KEK-TH-1893, KEK-Cosmo-192
}
\vspace{190mm}
\end{figure}

\begin{abstract}
In the past several decades, the standard cosmological model has been established and its parameters
have been measured to a high precision, while there are still many of the fundamental questions in cosmology;
such as the physics in the very early Universe, the origin of the cosmic acceleration and the nature of the dark matter.
The future world's largest radio telescope, Square Kilometre Array (SKA), will be able to open the new frontier of cosmology
and will be one of the most powerful tools for cosmology in the next decade.
The cosmological surveys conducted by the SKA would have the potential not only to
answer these fundamental questions but also deliver the precision cosmology.
In this article we briefly review the role of the SKA from the view point of the modern cosmology.
The cosmology science led by the SKA-Japan Consortium (SKA-JP) Cosmology Science Working Group
is also discussed.
\end{abstract}

\section{Introduction}

In the past several decades, by accurately measuring the statistical properties of the cosmic microwave background (CMB) anisotropies,
large-scale structures, supernovae and gravitational lensing, a simple standard cosmological model has been established and
its parameters have been measured to a high precision.
Our Universe is well described by a spatially flat, $\Lambda$ cold dark matter (CDM) model with nearly scale-invariant, 
adiabatic and Gaussian primordial fluctuations.
But the high-precision measurements have revealed that only $5\%$ in the energy budget of the Universe seems to be made
of the ordinary atomic matter \citep{Ade:2015xua}.
In other words, the remaining $95\%$ seems to be composed of exotic dark components such as dark matter and dark energy.

Even though the standard cosmological model has been established, there are still many fundamental questions in cosmology.
The remaining issues in cosmology would be as follow:
What type of the inflationary model is actually realized in the very early Universe?
What is the mechanism of generating primordial fluctuations?
What is the dark matter?
What is the physical origin of the cosmic acceleration, namely the dark energy or the modification of gravity theory? 

Although the CMB has been one of powerful observational tools for cosmology in the past decades, the next step towards
a precision cosmology will require extra information.
The future world's largest radio telescope, Square Kilometre Array (SKA)~\footnote{See http://www.skatelescope.org}, 
will provide the three-dimensional information by measuring the large scale structure, and will be one of the most powerful 
tools for cosmology and play an important role in addressing these issues.

The purpose of this document is to introduce Japanese scientific interests in the SKA project and to report results of our investigation. 
It is still in progress, so that the document may not fully cover previous works related to the SKA. We wish that the document 
becomes an interface for future communications, collaborations, and synergies with worldwide communities.

\subsection{Inflation}\label{sec:inflation}

Inflation, which leads to an epoch of exponential expansion in the very early Universe, 
was proposed as a solution 
to several problems in Big Bang cosmology, such as the horizon, flatness, and monopole problems.
Furthermore, it naturally sets the initial conditions required by the standard Big Bang cosmology.
A promising way to access the very early stage of the Universe 
is to measure the spatial pattern of the density fluctuation, because quantum fluctuations
of fields during the inflationary epoch can seed the primordial density fluctuations and 
different inflationary models predict different amplitude and scale-dependence of the primordial density fluctuations.
Hence observing density fluctuations provides us the rich information about not only
the evolution of the late-time Universe but also the primordial nature of the Universe, namely 
the inflation models as a mechanism for generating primordial fluctuations.

Although in the standard cosmology the primordial fluctuations are assumed to be Gaussian, 
recently possible small deviations from a purely Gaussian primordial density fluctuations, 
so-called {\it primordial non-Gaussianity}, have been investigated.
Since primordial non-Gaussianity reflects the fundamental interactions and nonlinear processes involved 
during and/or immediately after the inflation, it can bring insights into the fundamental physics behind inflation.
Hence this is one of the most informative fingerprints of inflation and is more generally key to understanding 
the extremely high-energy physics.
Among the several types of primordial non-Gaussianity, the local-type $f_{\rm NL}$ has been studied widely.
To quantify primordial non-Gaussianity of the local form we can parametrize
the Bardeen potential $\Phi$ with a random Gaussian field $\phi$ as~\citep{Gangui:1993tt,Komatsu:2001rj},
\begin{equation}
	\Phi =\phi +f_{\rm NL}\left(\phi^2 -\langle\phi^2\rangle\right)
	\,,
\end{equation}
where $\langle\cdots\rangle$ denotes the ensemble average.
If the distribution of density perturbations does not obey Gaussian statistics, we rather need
not only the power spectrum but also the higher-order statistics such as the bispectrum (three-point functions). 
It has been shown that even the simplest inflationary models predict small but nonvanishing deviation from Gaussianity; 
$f_{\rm NL}$ of ${\cal O}(0.01)$~\citep{Maldacena:2002vr}.
Primordial non-Gaussianity has primarily been constrained from the bispectrum 
in CMB temperature fluctuations and polarizations.
Recently, Planck \citep{Ade:2015ava} obtained a tight constraint of $f_{\rm NL}=0.8\pm 5.0$ at
$1\sigma$ statistical significance, implying that the inflationary models predicting too large
$f_{\rm NL}$ were already ruled out.
Although current CMB experiments are already close to cosmic-variance-limited ones, it is still rather weak
to distinguish the inflationary models.
A complementary way to access primordial non-Gaussianity is to measure its impact on large-scale structure.
While galaxy surveys are expected to constrain $f_{\rm NL}$ to the level comparable to CMB anisotropies,
the current constraint from the large-scale structure is about $\sigma (f_{\rm NL})\approx 50$ 
due to the unknown systematics~\citep{Ho:2013lda,Giannantonio:2013uqa}.

In addition to primordial density fluctuations, inflation predicts the existence of primordial gravitational waves, 
or tensor perturbations, while there is no evidence of the presence of primordial gravitational waves yet.
The odd-parity spatial pattern (B-modes) in the CMB polarization has been
considered as a prime observation to probe primordial gravitational waves~\citep{Kamionkowski:1996zd}.
The amplitude of tensor perturbations is conventionally parametrized by the tensor-to-scalar ratio, $r$, at a fiducial scale, and 
importantly the constraint on $r$ can be translated into the upper bound on the energy-scale of inflation.
The constraint on the tensor-to-scalar ratio inferred from the current CMB observations
is $r<0.07\,(95\%\,{\rm C.L.})$ \citep{Array:2015xqh,Ade:2015tva}, which corresponds to the upper bound of the inflationary energy scale of $10^{16}{\rm GeV}$.
Hence the detection and/or measurement of the signals generated by primordial gravitational waves 
offers an important clue to probe the extremely high-energy physics behind inflation.

While it is often claimed that inflationary models predict a spatially perfect flat geometry,
investigating the nonvanishing spatial curvature as an extension to the standard cosmology is still theoretically of interest.
The spatial curvature is not prohibited as the consequence of the inflation, and
indeed large amount of negative spatial curvature \citep{Gott:1982zf,Gott:1984ps} can be introduced 
in the open inflation scenario, namely the inflationary model with bubble nucleation by false-vacuum decay~\citep{Coleman:1977py,Coleman:1980aw}.
This type of models is recently attracting a renewed attention in the context of 
the string landscape paradigm~\citep{Susskind:2003kw,Freivogel:2004rd}.
Hence a search for the imprint of the false-vacuum decay in current/future observations such as CMB anisotropies 
is still very interesting. 
It is then shown that we are already testing the inflationary model with false-vacuum decay, which naturally predicts the spatially curved geometry,
against observations~\citep{Yamauchi:2011qq,Sugimura:2012kr}.

\subsection{Dark energy and modification of general relativity}
\label{sec:DE}

It is one of the biggest challenges of modern cosmology to understand
the physical origin of the cosmic acceleration of the Universe, discovered by observations of supernovae~\citep{Riess:1998cb,Perlmutter:1998np}.
It might eventually require the presence of a new type of matter, usually called dark energy.

The simplest model for dark energy is a cosmological constant, $\Lambda$, originally introduced by Einstein
a hundred years ago.
However, in this model there is a theoretical problem, the so-called cosmological constant problem.
Based on naive arguments in quantum field theory, the vacuum energy density 
due to the cosmological constant can be theoretically estimated from the sum of the contributions 
from quantum fields below an ultraviolet cutoff at the Planck scale.
On the other hand,
cosmologically observed value is many orders of magnitude smaller than this theoretical estimates by a factor of $10^{120}$.
So far it is still hard to naturally explain why such vanishingly small $\Lambda$ can be realized as a result
of a cancellation of more than $10^{120}$.
As the alternative form of the dark energy, quintessence, more precisely a scalar field rolling down its potential,
has been discussed~\citep{Wetterich:1987fm,Ratra:1987rm}. 
Unlike the cosmological constant, the quintessence is dynamic and the equation of state for the quintessence 
dark energy generally varies with time.

In another scenario, the accelerated expansion might arise alternatively due to a modification of general relativity 
on cosmological scales, which is often referred as {\it modified gravity theories}.
In addition to the standard tensor modes, this type of models often involves a new degree of freedom that accounts for cosmic acceleration.
Among them, the simple modification is described by a scalar-tensor theory where a single scalar degree of freedom is added.
A number of modified gravity theories consistent with current observations have been proposed so far as alternatives
to dark energy.
The $f(R)$ gravity \citep{DeFelice:2010aj}, DGP gravity \citep{Dvali:2000hr}, 
Galileons \citep{Nicolis:2008in}, and massive gravity \citep{Fierz:1939ix, deRham:2010kj}
(see also, e.g., \cite{Kimura:2013ika} for its extension) are possible candidates to explain 
the cosmic acceleration due to the additional degree of freedom.
Moreover, a very general class of modified gravity theories, called Horndeski theory \citep{Horndeski:1974wa} and its extension \citep{Gleyzes:2014dya, Gao:2014soa},
includes a large number of concrete models as specific cases.
Since modification of gravity are strongly constrained by precise tests of gravity on small scales such as one in the Solar
system, for a modified theory of gravity to be consistent, any of those involving an additional propagating
degree of freedom must include a mechanism that suppresses any potential fifth force that might appear on small scales.
A wide class of models can be shown to exhibit screening in the vicinity of a gravity source 
\citep{Kimura:2011dc,Kobayashi:2014ida}
and some observational implications were suggested in \citep{Narikawa:2013pjr,Saito:2015fza}.
A variety of theoretical scenarios have been proposed in literature and need to be carefully compared with observational data.

To elucidate the nature of the dark energy and  test gravity theory on cosmological scales,
we need to introduce parameters to describe non-standard cosmological models.
The simplest parametrization of background-level cosmological quantities is the equation-of-state (EoS) parameter,
which is defined as
\begin{equation}
	w=\frac{p}{\rho}
	\,,
\end{equation}
where $p$ and $\rho$ are the spatially averaged background dark energy pressure and energy density.
In the cosmological constant model, the EoS parameter is $w=-1$, but
if the dark energy is not a cosmological constant, $w\neq -1$, $w$ in general can vary in time.
In order to test a time-varying EoS, we for instance expand $w$ as
$w(z)=w_0+w_a\,z/(1+z)$.
The current constraints on $w_0$ and $w_a$ have been obtained from the combinations 
among data of CMB, supernovae, baryon acoustic oscillation and gravitational lensing and so on.
At the moment, there is no evidence for a departure from the standard $\Lambda$CDM model~\citep{Ade:2015rim}.
However, we should note that since there are vast possibilities to model the landscape of models, 
there is no general framework to parametrize the EoS parameter.
Hence more extensive studies of dark energy and modified gravity models are needed.
We expect that investigating the deeper Universe in future observations can help 
to break the degeneracy between dark energy and modified gravity.

\subsection{Dark matter, neutrino, and evolution of large-scale structure}\label{sec:DM}

Dark matter is crucial for the concordance cosmological model as a component which plays the central role in modeling
of cosmic structure formation and galaxy formation in the Universe.
The growth of the large-scale structure is rather sensitive to the properties of the dark matter and neutrinos.
Current observations on large scales are in good agreement with the prediction from the standard cosmological model, 
based on the cold dark matter, which is the matter composed of particles with the free-streaming length much smaller
than galaxy scales, and massless neutrinos.
However, the concordance model should yet be tested with observations on smaller scale. 

As a simple alternative to CDM, warm dark matter with its mass in the ${\rm keV}$ range has been proposed \citep{Dodelson:1993je,Markovic:2013iza}.
While above the specific scales charactering its mass the behavior of warm dark matter is very similar to CDM and is indistinguishable from CDM, 
it erases density perturbations and drastically suppresses the power spectrum below this scale through its free streaming.
On the other hand, the neutrino oscillation experiments by Super Kamiokande revealed that neutrinos must have small but 
nonvanishing masses, which indicates the existence of the physics beyond the standard model of the particle physics.
So far only mass-squared difference of neutrinos have been measured~\citep{Adamson:2008zt,Aharmim:2008kc}, 
suggesting that their masses are at least ${\cal O}(0.01){\rm eV}$. 
Massive neutrinos affect the evolution of the density perturbations and erases density perturbations through their free streaming.
Therefore, observing the small-scale power spectrum is of great interest not only for cosmology but also particle physics.


\section{Cosmological surveys with the SKA}\label{sec:surveys}

Observing the large-scale structure in the Universe provides us the rich information
about not only the late-time but also the primordial nature of the Universe.
However, a wide area of the sky and a significant redshift depth is essential
to achieve the precision at and beyond CMB levels.
Among the next-generation cosmological experiments, 
the SKA will be able to probe a vast volume of the Universe 
and is expected to deliver the very precise cosmological measurements 
and lead to the new frontier of cosmology science.
In this section, we briefly introduce major cosmology science partly developed by the international SKA cosmology
science working group, and partly presented in the international SKA Science Book (2015).

\begin{table*}
\caption{Specifications for SKA cosmology-oriented surveys.}
\vspace{10pt}
 \centering
 \begin{tabular}{c|l|ccccccc}\hline\hline
  survey & phase & coverage         & sensitivity           & resolution & obs. time & redshift & galaxy number & \\
           &          & (${\rm deg}^2$) &  ($\mu{\rm Jy}$) & (${\rm arcsec}$)        & (hr)        &           &         & \\ 
  \hline
  \multirow{2}{*}{HI galaxy}
  & SKA1 MID(B1+2) & $5,000$   & $\simeq 70$ & $\lesssim 1$ & $10^4$ & $z\lesssim 0.7$ & $\simeq 5\times 10^6$ \\\cline{2-9}
  & SKA2            & $30,000$  & $\simeq 5$    & $\lesssim 0.1$ & $10^4$ & $z\lesssim 2$   & $\simeq 10^9$ & \\\cline{2-9}
  \hline
  \multirow{2}{*}{HI IM}
  & SKA1 MID(B1+2)/LOW & $30,000$  & --    & $\lesssim 1/\lesssim 10$ & $10^4$ & $z\lesssim 3$   & -- \\\cline{2-9}
  & SKA2                       & $30,000$  & --    & $\lesssim 0.1$ & $10^4$ & $z\lesssim 3.7$   & -- & \\\cline{2-9}
  \hline
  \multirow{2}{*}{continuum}
  & SKA1 MID(B2) & $30,000$ & $\simeq 1$    & $\lesssim 1$ & $10^4$ & $z\lesssim 6$   & $\simeq 10^8$ \\\cline{2-9}
  & SKA2            & $30,000$ & $\simeq 0.1$  & $\lesssim 0.1$ & $10^4$ & $z\lesssim 6$   & $\simeq 10^9$ & \\\cline{2-9}
  \hline\hline
\end{tabular}
\label{table:survey design}
\end{table*}

\subsection{HI galaxy survey}

Neutral hydrogen (HI) in the Universe can be a prime candidate to observe as the tracer 
of the underlying dark matter and has a characteristic line emission whose wavelength is given by $21$cm. 
Since after the Universe has reionized, the HI is thought to mainly live inside galaxies, the HI is therefore
essentially a biased tracer of the galaxy distributions~\citep{Yahya:2014yva}.
The redshifting of HI ($21$cm) line provides the redshift information of galaxies and helps to reconstruct
the three-dimensional matter distributions over a range of redshifts and scales. 
This is called {\it HI galaxy redshift survey}.
However, the line is quite weak, we require highly sensitive telescopes such as the SKA to perform
large HI galaxy survey across a wide redshift range, while 
so far only galaxies up to $z\sim 0.2$ have been detected in Arecibo~\citep{2015MNRAS.446.3526C}.

Detection of a HI galaxy relies on the measurement of its corresponding HI line profile.
The choice of source detection algorithm for the HI galaxy surveys is crucial and determines the total number of galaxies 
and how well spurious detections, e.g., due to radio frequency interference (RFI), can be rejected.
In the simple approach, which has been often studied (e.g., see \citep{Abdalla:2015kra,Santos:2015hra,Yahya:2014yva}), 
we require that at least two points on the HI line are measured. 
This is because expected HI galaxies show the typical signals of the double peak due to their rotation and 
this method will remove any face-on galaxy showing a single narrow peak, which could be confused with RFI. 

Based on this method, we expect that
the HI galaxy redshift survey provided by SKA has the potential to be 
competitive with other cosmological experiments in the next decade~\citep{Abdalla:2015kra,Santos:2015hra,Yahya:2014yva}.
Assuming $10^4$ hours of survey time, the SKA phase 1 (SKA1) covers $5,000\,{\rm deg}^2$ out to $z\sim 0.5$, 
while SKA phase 2 (SKA2) should probe the available full-sky, $30,000\,{\rm deg}^2$ up to $z\sim 2$.
For SKA1 with flux sensitivity $70-100\mu$Jy, we will find about $5\times 10^6$ HI galaxies,
which is slightly smaller than one expected in planed galaxy surveys in optical and infrared band,
such as Euclid~\footnote{See http://www.euclid-ec.org} and LSST~\footnote{http://www.lsst.org}.
For comparison, the SKA precursors, MeerKAT and ASKAP, with the sky coverage $5,000\,{\rm deg}^2$ and 
flux density $\sim 700\,\mu$Jy are expected to find $5\times 10^5$ galaxies.
When the SKA2 is constructed, the flux threshold will be drastically improved and can reach $\sim 5\mu$Jy
over the available full-sky, providing that the spectroscopic survey of $\sim 1$ billion HI galaxies,
that is, ``{\it billion galaxy survey}'', can be delivered.
Yahya et al.\,(2015)
shows that the number density and the bias for the HI galaxy survey
conducted by the SKA can be well described by the fitting function with several parameters depending on the flux cut.
Therefore, the SKA cosmology survey will provide an unprecedented number of galaxies, compared with
any planned galaxy surveys in 2020s.

The SKA HI galaxy survey in its both phase will be able to provide very accurate measurements for
the expansion history of the Universe via baryon acoustic oscillation (BAO)~\citep{Bull:2015nra},
which is imprinted in the galaxy correlation function as the standard ruler, 
and the growth of structure via redshift space distortion (RSD)~\citep{Raccanelli:2015qqa}.
It is shown that the growth rate of the observed density contrast is sensitive to the dark energy models 
and the theory of gravity (see Sec.~\ref{sec:DE}).
Hence the precise RSD measurement then provides a test for deviations from general relativity on large scales.

\subsection{HI intensity mapping survey}

In addition to the HI galaxy redshift survey, the SKA will be also able to deliver competitive cosmology
by the {\it HI intensity mapping (IM) survey} \citep{Santos:2015gra}
(see also \cite{Battye:2004re,Wyithe:2007gz,Chang:2007xk,Peterson:2009ka}), 
which offers novel technique of probing the large-scale structure of the Universe.
For the HI galaxy survey, in order to detect individual galaxies, 
we indeed need sufficiently long integration time of the highly sensitive telescope.
Instead of resolving individual galaxies, the HI IM survey can measure 
the intensity of the integrated HI line emission of several galaxies in one angular pixel on the sky.
For the SKA, the pixel size is usually assumed to be $\lesssim 1\,{\rm deg}^2$~\citep{Santos:2015gra}.
Similar to the CMB map, the result of the HI IM survey is a map of large-scale 
fluctuations in HI intensity.
By not requiring the detection of individual galaxies and combining with the high resolution
of radio telescope, even for the SKA1, the HI IM survey can cover the extremely 
large survey volume that is larger than that of future planed ones.
When the SKA1-MID IM survey operating for $10^4$ hours with the bandwidth $350-1050\,{\rm MHz}$ is considered,
the SKA1 can probe the available full-sky $30,000\,{\rm deg}^2$ out to $z\sim 3$.
We note that pushing to higher redshifts brings in issues of angular resolution and foreground, 
while the detection of IM at lowest redshifts is known to be difficult due to the RFI~\citep{Bull:2014rha}.
When the whole sample is assumed to be further subdivided into constant frequency bins of $\sim 10\,{\rm MHz}$ width,
more than $50$ redshift bins can be used.

Even though individual galaxies are not detected, the HI IM survey has
a sufficient potential to deliver precise measurements of the BAO~\citep{Bull:2015nra} and RSD~\citep{Raccanelli:2015qqa}
in the matter power spectrum even in the era of SKA1.
These surveys provide powerful tests of dark energy and modifications to general relativity,
which are discussed in Sec.~\ref{sec:DE}.
Moreover, the very large survey volume provided by the HI IM survey provides us 
the possibility to probe the physics on extremely large-scales~\citep{Camera:2015yqa},
which carries the information of the possible signature of the primordial inflation in the very early Universe 
(see Sec.~\ref{sec:inflation}).
In addition to these, the HI IM survey could allow the measurements of the power spectrum for the lensing potential 
at higher redshifts~\citep{Brown:2015ucq} (see also \cite{Pourtsidou:2013hea}),
mainly because gravitational magnification due to weak gravitational lensing affects the clustering properties of galaxies. 
The radio weak lensing survey would map the dark matter distributions (see Sec.\ref{sec:DM}) 
in the independent way from optical counterparts.

Observing the HI signals emitted at the Epoch of Reionization (EoR) can also be used to deliver
the very precise cosmology.
During the EoR, the first stars and galaxies release ionizing photons into the intergalactic medium (IGM),
creating ionized bubbles whose geometry reflects the large-scale clustering of the dark matter.
Since the HI gas survives not only inside galaxies in the ionized bubble but also in the neutral IGM
during the EoR, a survey of the HI brightness temperature fluctuations can be treated as another biased tracer of
the underlying dark matter distribution~\citep{Santos:2015gra}, as with galaxy surveys.
In addition, the HI IM survey before and during EoR can also be 
used to measure the weak gravitational lensing of the redshifted HI emissions 
to reconstruct the high-redshift matter density fluctuations along the line-of-sight~\citep{Zahn:2005ap}.

Since the astrophysical processes during the EoR are poorly understood and they lead to large
uncertainties, a number of parameters to model the systematics should be included and 
the marginalization of the contributions from the astrophysical process is needed.
Furthermore, the HI signals are heavily contaminated by various sources such as our galaxy and
extragalactic point sources. 
One of the important challenges for the HI IM survey is to develop
the cleaning method to remove the foregrounds
that are orders of magnitude stronger than the signals to be measured.
Although many works concerning the cleaning techniques have already been done, e.g., 
\citep{Wolz:2015sqa,Shaw:2013wza,Chapman:2012yj,Santos:2004ju}
(see \cite{Ichiki:2013cva} for the CMB foreground removal), 
it is still challenging to remove the everything that is not what we actually want to observe.

\subsection{Radio continuum survey}\label{sec:RC}

Although in the past decades the {\it radio continuum survey} has provided 
distribution of the large scale structure in low redshift, 
the studies from cosmological point-of-view has been
restricted, e.g. the cross-correlations with the CMB to find the integrated Sachs-Wolfe effect (e.g., \cite{Giannantonio:2008zi}), 
mainly because the source density in radio continuum survey is relatively lower than that in ongoing optical surveys.
When the SKA is constructed, the radio continuum survey will be expected to provide a sufficient number of sources
to be comparable to or relatively larger than expected in forthcoming surveys in other wavelength~\citep{Jarvis:2015tqa}.
Synchrotron radiation is emitted from all galaxies with ongoing star formation or accretion
and dominates the extragalactic radio background emission in low frequency regime.
Moreover, since the synchrotron radio emission from galaxies is known to be unaffected by dust, 
the radio continuum survey is advantageous to detect high-redshift galaxies.
Because the semi-empirical extragalactic sky simulation performed in \citep{Wilman:2008ew} is so far in good
agreement with latest radio observations such as JVLA~\citep{Condon:2012ug},
the resultant redshift distributions of various source populations with several detection thresholds 
can be used to forecast future constraints on cosmological parameters.
In addition, the radio continuum survey from SKA will cover the available full-sky $30,000\,{\rm deg}^2$
out to $z\sim 6$, which can probe extremely large survey volume.
On the other hand, since the radio continuum survey provides the featureless spectrum, the redshift
information for individual galaxies is not available and can be obtained from other wavelength
observations or HI line surveys.
For instance, using observational data from optical surveys, the host galaxies of low-redshift radio sources will 
be expected to be identified, while radio sources at high redshifts, say $z>2$, will not be identified by optical surveys~\citep{McAlpine:2012cu}.

The SKA1 has the potential to perform a radio continuum survey over the available full-sky 
down to the flux threshold $1\,\mu{\rm Jy}$, which corresponds to the detection of $\sim 5\times 10^8$ galaxies.
The SKA2 with the flux limit $0.1\mu{\rm Jy}$ can reach $10^9$ galaxies.
In the absence of redshift information, the continuum survey mainly observes the angular power spectrum for
the number density contrast of radio sources.
Even without redshift information for individual galaxies, the deep and wide surveys are advantageous
to constrain the information of the primordial Universe, e.g., the primordial non-Gaussianity (see Sec.~\ref{sec:inflation}),
as shown in the following section.

The radio continuum survey will also measure the deformation of the distant-galaxy images, 
providing the high-precision measurements of cosmic shear~\citep{Brown:2015ucq}.
It directly offers an opportunity to probe intervening total matter fluctuations along a line-of-sight.
The cosmic shear measurement of background source has been widely studied and now been accepted
as a standard cosmological tool.
The SKA1 will survey $5,000\,{\rm deg}^2$ and the SKA2 can reach $30,000\,{\rm deg}^2$.
We will find galaxy number densities of about $3$ galaxies per arcmin$^2$ for SKA1
and about $10$ galaxies per arcmin$^2$ for SKA2. 
While the number of galaxies is relatively lower,
the radio continuum survey can extend to higher redshift than the optical weak lensing surveys
that will be conducted on comparable timescales.


\section{Cosmology with the SKA by SKA-Japan consortium}

As introduced in the previous section, the SKA has the potential to deliver
the very precise cosmological measurements through the HI galaxy survey, 
HI intensity mapping survey and radio continuum survey. 
We expect that the cosmological survey conducted by the SKA will yield transformational science 
across a wide range of cosmology.
In this section, we summarize the SKA-oriented cosmological scientific challenges in which 
the cosmologists in Japanese research community, called SKA-Japan consortium (SKA-JP)
cosmology science working group, have a deep interest.
We have wide expertise in both theoretical/observational cosmologies
and are particularly interested in the studies of the very early stage
of the Universe, the observational tests of the general relativity on cosmological scales and 
the particle physics such as the neutrinos and the beyond-standard-model physics.
While observing the small-scale is of great interest from various perspectives for both cosmology and particle physics, 
it is hard to provide the accurate predictions on such scales, mainly because density perturbations
grow significantly and enter within highly nonlinear regime.
Moreover, the feedback of the baryonic physics is poorly understood and involves large uncertainties.
In order to obtain the unprecedented constraints on the parameters for the fundamental physics, 
we need to focus on the observation on the specific redshifts and scales.
This section is a brief overview of the science that SKA can achieve and our community
is currently playing an important role.

\subsection{Ultra-large scale cosmology with multitracer technique}\label{sec:multitracer technique}

One of the possible ways to overcome the issues discussed above is to focus on the observations on the ultra-large 
cosmological scales.
Although it is hard to access ultra-large scales, theoretical prediction on such scales
can be very precise, because in superhorizon scale, the growth of the density perturbation
fully remains within the linear regime and the effects of the baryonic physics feedback 
due to the astrophysical processes are sufficiently suppressed.
However, the clustering analysis at large scales are limited due to cosmic-variance,
due to the lack of enough independent measurements.
A promising way to access the ultra-large scales is to reduce the cosmic-variance with
so-called multitracer technique~\citep{Seljak:2008xr,Hamaus:2011dq}, in which the availability of multiple
tracers with the different biases allows significant improvements in the statistical error.

\subsubsection{Multitracer technique and synergy with Euclid survey}

\begin{figure}
 	\begin{center}
		\includegraphics[width=8cm]{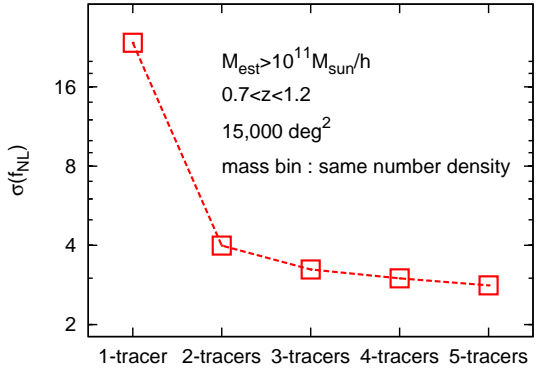} 
	\end{center}
	\caption{
		The marginalized error on $f_{\rm NL}$ as the function of the number of tracers
		in the single redshift bin $0.7<z<1.2$. The mass bins are divided such that they
		have the equal shot noise.
		Even $2$ tracers drastically improve the constraint, because the multitracer
		technique does not take effect for the one tracer case. \citep{Yamauchi:2014ioa}
	}\label{fig:sigma_fNL_nTracer}
\end{figure}

Among various cosmological parameters characterizing the primordial Universe,
we focus on the primordial non-Gaussianity.
Many members of Japanese cosmology community
have been extensively working on the inflationary model-building and the theoretical studies of the mechanism
of generating the primordial non-Gaussianity during the inflationary epoch.
Although primordial non-Gaussianity has primarily been constrained from measurements of CMB so far, 
the resultant constraint has almost saturated to the observational limit predicted by ideal observations.
Large scale halo/galaxy distributions provide distinct information on the primordial non-Gaussianity.
Luminous sources such as galaxies must be most obvious tracers of the underlying dark matter distributions with a bias.
For the standard Gaussian initial conditions, the halo bias is often assumed to be linear, deterministic, and scale-independent. 
However, it was found that the non-Gaussianity in primordial fluctuations effectively introduces 
a {\it scale-dependent} clustering of galaxies on large scales.
The modification appears in a biased tracer of the underlying matter distribution.
The non-Gaussian correction to the large-scale Gaussian bias of a biased tracer, 
induced by the local-type non-Gaussianity $f_{\rm NL}$, is given by~\citep{Dalal:2007cu,Matarrese:2008nc}.
\begin{equation}
	\Delta b=\frac{3H_0^2\Omega_{{\rm m},0}\delta_{\rm c}(b_{\rm G}-1)}{k^2T(k)D_+(z)}f_{\rm NL}
		-\frac{1}{\delta_{\rm c}}\frac{1}{{\rm d}\ln\nu}\left(\frac{{\rm d}n/{\rm d}M}{{\rm d}n_{\rm G}/{\rm d}M}\right)
	\,,
\end{equation}
where $\nu =\delta_{\rm c}/\sigma (M)$, 
$D_+(z)$ is the growth factor and $T(k)$ is the matter transfer function normalized to unity at large scales.
The effect of the local type primordial non-Gaussianity on the halo bias is prominent on large scales
and at high redshifts.
Thus, in the large-scale limit, the non-Gaussian correction become dominant, and the enhancement of clustering amplitude
becomes significant.
With the help of this property, the constraint on primordial non-Gaussianity has been obtained
(see e.g., \cite{Alonso:2015uua,Camera:2014bwa}).

\begin{figure}
 	\begin{center}
		\includegraphics[width=8cm]{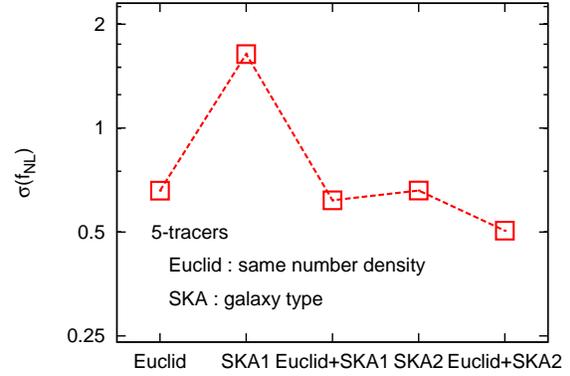} 
	\end{center}
	\caption{
		The expected marginalized constraints on $f_{\rm NL}$ for each survey and combinations. 
		Constraints $\sigma (f_{\rm NL})\lesssim 1.8$ and $0.7$ can be obtained with a single survey of SKA1 and Euclid, respectively.
		The combination of SKA and Euclid can provide a stronger constraint $\sigma (f_{\rm NL})\lesssim 0.65$
		for SKA1+Euclid and an unprecedented constraint $\sigma (f_{\rm NL})\lesssim 0.5$ for SKA2+Euclid.~\citep{Yamauchi:2014ioa}
	}\label{fig:sigma_fNL_survey}
\end{figure}

D.~Yamauchi and K.~Takahashi, who are members of SKA-JP, have discussed the potential power of the multitracer technique with
future galaxy surveys in Yamauchi et al.\,(2014) (see also \cite{Ferramacho:2014pua}). 
Hereafter we assume that the stochastic part of the halo clustering can be approximately described
by the diagonal shot noise given by one over the number density, while the numerical simulation~\citep{Hamaus:2010im} has 
suggested small deviations from this description.
Before showing expected constraints from SKA, the specific galaxy survey with single redshift bin $0.7<z<1.2$, 
the sky coverage $15,000\,{\rm deg}^2$ and the observed halo mass range $M_{\rm est}>10^{11}M_\odot /h$ 
is considered to investigate the efficiency of the multitracer technique.
We found that the constraints on $f_{\rm NL}$ drastically improved even when a number of nuisance parameters for estimates 
of the halo mass are included to model systematic errors （see Fig.~\ref{fig:sigma_fNL_nTracer}）.
Even for the case of $2$ tracers, the constraint of $\sigma (f_{\rm NL})\approx 4$, which is severer than the Planck one
$\sigma (f_{\rm NL})\approx 5$, can be obtained.

We then focus on the SKA radio continuum survey (Sec.~\ref{sec:RC}).
In our treatment, the specific population of galaxies
is considered to match the extragalactic simulation performed in Wilman et al.\,(2008). 
assuming that the sensitivity limit of $1\,(0.1)\mu{\rm Jy}$ and applying a $5\sigma$ detection for SKA1(2).
In this case, they correspond to the detection of $5\times 10^8$ and $10^9$ galaxies, respectively.
Moreover, to split the galaxy samples into the subsamples by the inferred halo mass and redshift, 
five radio galaxy types such as star-forming galaxies, radio quite quasars, radio-loud AGN and starbursts are 
considered to infer the halo mass.
We actually have shown that the SKA radio continuum survey can constrain up to the accuracy of 
$\sigma (f_{\rm NL})\approx 0.7$ for SKA2, and $\sigma (f_{\rm NL})\approx 1.6$ for SKA1.
Even when the GR horizon-scale effects are taken into account, it is shown that $\sigma (f_{\rm NL})\lesssim 1$
can be achieved~\citep{Alonso:2015sfa,Fonseca:2015laa}.

At the time when the SKA is operational, there will be additional survey from a number of ground-based telescopes
and space missions.
Euclid is expected to be launched in 2020 and will perform
imaging and spectroscopic surveys in optical and infrared bands.
While the number of galaxies ($\sim 10^8$)
and covered sky area ($\sim 15,000\,{\rm deg}^2$)
for Euclid are relatively smaller than these for SKA,
we consider Euclid photometric galaxy survey as a future representative survey that will be conducted on comparable time scales.
Fig.~\ref{fig:sigma_fNL_survey} shows that constraints $\sigma (f_{\rm NL})\lesssim 1.8$ and $\lesssim 0.7$ can be obtained with a single survey
of SKA1 and Euclid, respectively.
The combination of SKA and Euclid can provide a stronger constraint $\sigma (f_{\rm NL})\lesssim 0.65$ for SKA1+Euclid
and an unprecedented constraint $\sigma (f_{\rm NL})\lesssim 0.5$ for SKA2+Euclid,
implying that most inflationary models with even small primordial non-Gaussianity 
can be excluded by the combination of future galaxy surveys (see also \cite{Kitching:2015fra,Takahashi:2015zqa}).

\subsubsection{Higher-order primordial non-Gaussianity and their consistency relation}

\begin{figure}
 	\begin{center}
		\includegraphics[width=8.5cm]{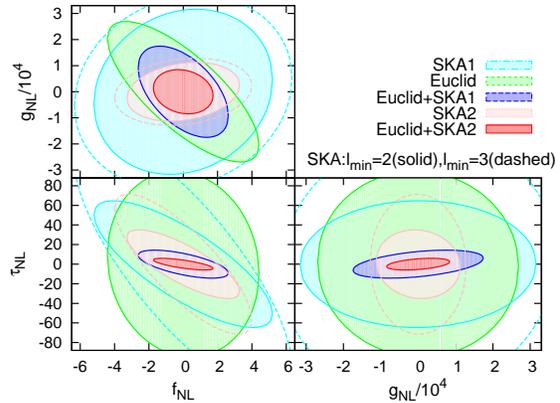} 
	\end{center}
	\caption{
		Forecast $1\sigma$ marginalized contours in $(f_{\rm NL},\tau_{\rm NL})$,
		$(g_{\rm NL},\tau_{\rm NL})$ and $(f_{\rm NL},g_{\rm NL})$ planes with
		the vanishing fiducial values of the nonlinear parameters. 
		To see the dependence on the minimal multipole we also plot the results for SKA
		with $\ell_{\rm min}^{\rm SKA}=3$ in dashed line. \citep{Yamauchi:2015mja}
	}\label{fig:fNL-tauNL-gNL_Fisher}
\end{figure}

Constraining various properties of primordial non-Gaussianity is expected to provide the opportunity 
to break the degeneracy of inflationary models, because its amplitude, shape, 
and scale-dependence encode much information of the primordial Universe.
In other words, different nonlinear parameters are linked to distinctive mechanisms that can generate
such non-Gaussianity during/after the inflationary epoch.

One of the major theoretical discoveries from the studies of inflationary models
is that {\it all} inflationary models predict
the presence of the consistency relation between nonlinear parameters.
For the simplest scenarios, if there is the nonvanishing local-form bispectrum, 
the trispectrum must necessarily exist with
$\tau_{\rm NL}=((6/5)f_{\rm NL})^2$, where $\tau_{\rm NL}$ denotes the local-type nonlinear 
parameter charactering the amplitude of the primordial trispectrum~\citep{Okamoto:2002ik,Boubekeur:2005fj}.
Even in a general situation, one can show that there is a universal relation, 
\begin{equation}
	\tau_{\rm NL}\geq \left(\frac{6}{5}f_{\rm NL}\right)^2
	\,,
\end{equation}
which is often referred to the Suyama-Yamaguchi inequality~\citep{Suyama:2007bg,Suyama:2010uj}.
Primordial trispectrum is also usually characterized by another nonlinear parameter, $g_{\rm NL}$, 
which corresponds to the strength of the intrinsic cubic nonlinearities of primordial fluctuations.
A detection of the higher-order primordial non-Gaussianity and the confirmation of the inequality 
from large-scale structure measurements 
would indicate the presence of more complicated dynamics, e.g., multi-field inflationary models
(see \cite{Lyth:2001nq,Moroi:2001ct,Enqvist:2001zp} for curvaton model), 
in the primordial Universe.
Thus, the detection and confirmation should be the target in future experiments (see \cite{Biagetti:2012xy}).
Furthermore, we could rule out not only the simplest single-field but also the multifield inflation models 
if we find the breaking of the consistency inequality.

\begin{figure}
 	\begin{center}
		\includegraphics[width=8.5cm]{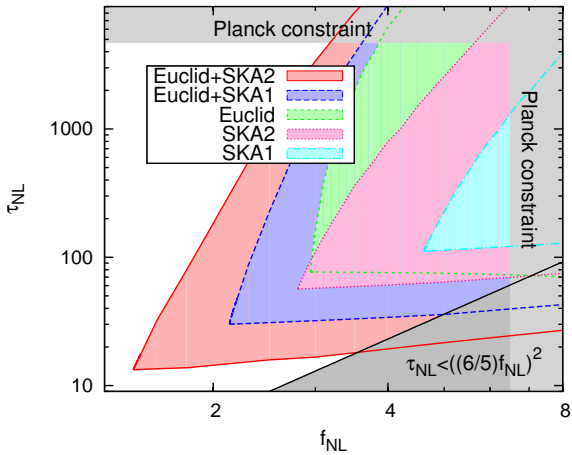} 
	\end{center}
	\caption{
		Parameter space to confirm the consistency relation,
		$f_{\rm NL}/\sigma (f_{\rm NL})$ and $\tau_{\rm NL}/\sigma (\tau_{\rm NL})$.
		For comparison, the inconsistent region and the constraints from Planck are
		also shown in gray color. 
		When $f_{\rm NL}\lesssim 1.5$, the confirmation of the consistency inequality is
		rather challenging for a single survey.
		Combining SKA2 and Euclid drastically improves the constraint to reach
		the confirmation even for $f_{\rm NL}\approx 0.9$ and $\tau_{\rm NL}\approx 8$.
		\citep{Yamauchi:2015mja}
	}\label{fig:sig}
\end{figure}

When $\tau_{\rm NL}>((6/5)f_{\rm NL})^2$, the halo density contrast is not 
fully correlated to the matter linear density field even in the absence of the shot noise.
Namely, the simple bias relation $\delta_{\rm h}=b_{\rm h}\delta$ should be modified due to
the stochasticity.
This provides the unique opportunity of large-scale scale-dependent {\it stochastic}
bias as a probe of the primordial Universe associated with complicated 
dynamics~\citep{Baumann:2012bc,Smith:2010gx,Adhikari:2014xua} .

D.~Yamauchi and K.~Takahashi have discussed the required survey level needed to
probe the higher-order primordial non-Gaussianity such as $\tau_{\rm NL}\,,g_{\rm NL}$ and 
test the consistency relation with future galaxy surveys~\citep{Yamauchi:2015mja} (see also \cite{Hashimoto:2015tnv} ).
In order to see the impact of the higher-order primordial non-Gaussianity on the parameter estimation, 
the $1\sigma$ marginalized contour obtained through the Fisher analysis is shown in Fig.~\ref{fig:fNL-tauNL-gNL_Fisher}.
Assuming the specific distribution of galaxies introduced in the previous subsubsection, as for $\tau_{\rm NL}$, 
the SKA radio continuum survey can reach $\sigma (\tau_{\rm NL})\approx 23\,(43)$ for the SKA2(1), 
which is an improvement by a factor of $100$ compared with the Planck constraint,
suggesting that the deep survey and large sky coverage provided by SKA are advantageous to constrain $\tau_{\rm NL}$, and
the information even from the early stage of SKA is quite essential to break the degeneracy
between these nonlinear parameters.
The constraint on $g_{\rm NL}$ from SKA is also several tens of times severer than
that obtained from Planck. 
Concentrating on $f_{\rm NL}$ and $\tau_{\rm NL}$, 
the dependence of our forecast on the choice of the fiducial values is studied in  Fig.~\ref{fig:sig},
where we show the region in which both $f_{\rm NL}$ and $\tau_{\rm NL}$ are detected at $1\sigma$ significance.
As is anticipated, the confirmation of the inequality becomes harder as decreasing $f_{\rm NL}$.
On the other hand, with increasing $\tau_{\rm NL}$, the constraining power on $f_{\rm NL}$ decreases,
mainly because the correction from $\tau_{\rm NL}$ to the halo bias dominates that from $f_{\rm NL}$.
This implies that the relatively smaller $\tau_{\rm NL}$ is needed to detect $f_{\rm NL}$.
Hence there is the wedge-shaped region where we can confirm the consistency inequality at more than $1\sigma$ level.
When $f_{\rm NL}$ is small, say $\lesssim 2.7$, the confirmation at $\gtrsim 1\sigma$ level is rather challenging
for a single survey, even when we apply the multitracer technique.
However, the combination of SKA and Euclid drastically improves the constraints and can detect the consistency 
relation in the wide parameter region.
These results shows that the SKA radio continuum survey combined with the future optical galaxy 
surveys such as Euclid offers an unique opportunity to probe the detail dynamics 
in the primordial Universe.

\subsubsection{Halo/galaxy bispectrum}

The primordial bispectrum and the higher-order moments measure 
the fundamental interactions and nonlinear processes occurring during and/or immediately after inflation.
Not only the amplitude but also the shape of the bispectrum encodes much physical
information of the primordial Universe. 
Different shape can be also linked to different mechanism for generating non-Gaussian fluctuations.
It would be very interesting to try further constraining various types of primordial non-Gaussianity
by precise large-scale structure measurements, 
which are expected to contain much more information than CMB.
It is, however, shown that the scale-dependent clustering due to non-local type models is weaker than 
the local-type~\citep{Matsubara:2012nc,Schmidt:2010gw,Verde:2009hy,Taruya:2008pg},
implying that the measurements of the scale-dependent clustering in the galaxy power spectrum is irrelevant.
The effect of the scale-dependent bias sourced by primordial non-Gaussianity appears not only
in the galaxy power spectrum but also in the galaxy bispectrum~\citep{Yokoyama:2013mta}.
Indeed, the amplitude of the galaxy bispectrum sourced by the equilateral-type primordial bispectrum and trispectrum,
which do not lead to an enhancement in the galaxy power spectrum, is shown to 
be enhanced on large scales~\citep{Sefusatti:2007ih,Mizuno:2015qma}.
Although the late-time nonlinear gravitational evolution effectively gives the nonvanishing contributions 
to the observed galaxy bispectrum, the different scale-dependence on large scales is expected to
give stringent constraints on the non-local type primordial non-Gaussianity.
Hence our community will try to constrain the non-local type primordial non-Gaussianity with the help of
the multitracer technique (see \cite{Yamauchi:2016wuc}).

\subsubsection{Relativistic effect on large scales}

Although the standard analysis of current observations is based on the Newtonian,
for the SKA covering the huge survey volume and the deep redshift depth
the Newtonian analysis is inadequate, because we will measure the correlation on scales 
above the Hubble horizon.
Therefore, to incorporate the superhorizon scales consistently the Newtonian analysis must be
replaced by the general relativistic one.
In order to map the distributions of the matter density field even in the ultralarge scales,
we need to correctly identify the galaxy overdensity observed on the past light cone.
Two explicit examples of such effects are RSD and gravitational lensing magnification, which
are expected to give the largest contributions to the observed galaxy overdensity.

The general relativistic analysis of the power spectrum has recently been developed in \cite{Yoo:2010ni,Bonvin:2011bg,Challinor:2011bk}, 
and the detectability of these effects by future surveys, as well as measurements 
of primordial non-Gaussianity, has been discussed in \cite{Alonso:2015sfa,Fonseca:2015laa}.
On the other hand, as mentioned in Sec.~\ref{sec:DE}, the late-time cosmic acceleration might arise
the modification of general relativity on cosmological scales.
If the gravity theory on superhorizon scales cannot be described by the general relativity, 
the resultant relativistic effect on the observed galaxy overdensity should be replaced by the correct one.
Hence we will try to derive the explicit form of the relativistic correction for the galaxy overdensity, 
in the context of the modified gravity theory.

\subsection{Exploring the dark Universe with $21$-cm survey}\label{sec:dark Universe}

%
As we saw in the Sec.~2.2,
in recent years,
observations of HI signals (i.e. 21 cm line) emitted at the EoR
have attracted much attention  
because the observations will open a new window to the early phases of cosmological structure formations.
%
%
Since there are a lot of hydrogen gas in the IGM
at the very high redshift Universe ($z\gtrsim 6$),
by observing the redshifted 21 cm line with SKA-low
we can survey such very past epochs,
and not only study how the Universe was ionized
but also obtain information about density fluctuations of matters.

The 21 cm line results from  hyperfine
splitting of the 1S ground state due to an interaction of  magnetic moments of proton and  electron,
and we can define the spin temperature $T_{S}$ 
through a ratio between  number densities of
hydrogen atom in the 1S triplet and 1S singlet states,
$n_{1}/n_{0}\equiv \left(g_{1}/g_{0} \right)\exp(-T_{\star}/T_{S})$,
where $T_{\star}\equiv hc/k_{B}\lambda_{21} = 0.068$~K with
$\lambda_{21} (\simeq 21$ cm) being the wave length of the 21 cm line
at a rest frame, and  $g_{1}/g_{0}=3$ is the ratio of  spin degeneracy
factors of the two states.  A difference between the observed 21 cm
line brightness temperature at redshift $z$  and the CMB temperature
$T_{{\rm CMB}}$ is given by
\begin{eqnarray}
\Delta T_{b}\left(\mbox{\boldmath $r$},z \right)  
& \simeq & \ 26.8 
x_{{\rm HI}}(1+\delta_{b})
         \left( \frac{\Omega_{b}h^{2} }{0.023} \right)
         \left( \frac{0.15}{\Omega_{m}h^{2}}\frac{1+z}{10} \right)^{1/2} \nonumber \\
       & & \hspace{-35pt} \times \left( \frac{1-Y_{p}}{1-0.25} \right) \left( \frac{T_{S}-T_{{\rm CMB}}}{T_{S}} \right)
         \left( \frac{H(z)/(1+z)}{dv_{\parallel}/dr_{\parallel}} \right) \ {\rm mK},
\end{eqnarray}
where $\mbox{\boldmath $r$}$ is the comoving coordinate,
$x_{{\rm HI}}$ is the neutral fraction of hydrogen, 
$Y_{p}$ is the primordial $^4$He mass fraction,
$\delta_{b}$ is the hydrogen density fluctuation, and $dv_{\parallel}/dr_{\parallel}$
is the gradient of the proper velocity along the line of sight due to
both the Hubble expansion and the peculiar velocity.
Through the contribution of $\delta_{b}$ in the differential brightness temperature of 21 cm line,
we can obtain the information of matter density fluctuations.

The growth of the density fluctuations depends on cosmological parameters,
and we can use the 21 cm line surveys for constraining them
like CMB observations or galaxy surveys.
Since  the redshifted frequency of 21 cm line corresponds to each past era,
the 21 cm line surveys enable us to observe very wide redshift ranges (21 cm tomography).
Therefore, we can obtain a large number of Fourier samples 
and information about time evolution of density fluctuations.
Additionally, in such high redshift eras, 
non-linear growth of density fluctuations is less effective
in comparison with that in later epochs,
and we can treat their time evolution with linear theory up to relatively small scales. 
For this reason, in the 21 cm line surveys,
theoretical uncertainties due to the growth of density fluctuations
are much smaller than those in galaxy surveys.
These features are  advantages of the 21 cm line surveys.

\begin{figure}[t]
  \begin{center}
   \includegraphics[width=8cm]{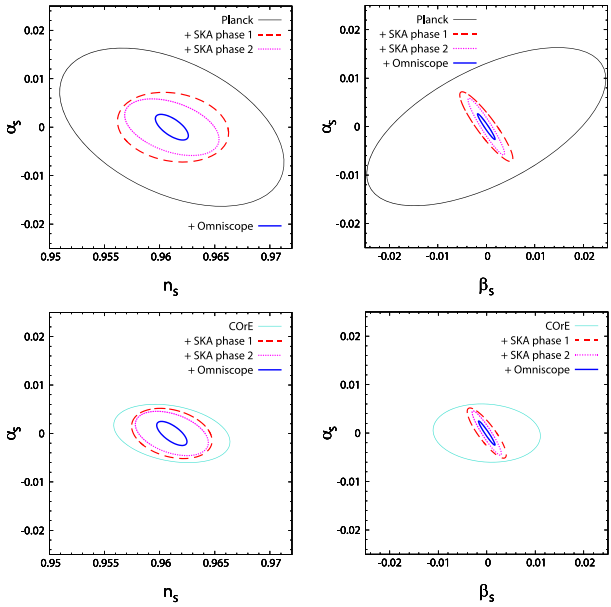} 
  \end{center}
\caption{
Contours of 95\% C.L. forecasts by
combinations of CMB (Planck (upper) and COrE (lower)) and SKA (phase~1 and phase~2)
in the $n_s$--$\alpha_s$ (left) and $\alpha_s$--$\beta_s$ (right) planes.
We set the observed redshift range to be $6.75<z<10.05$.
SKA can strongly constraint on $n_{s}, \alpha_{s}$ and $\beta_{s}$
in comparison with cases of CMB only.
For comparison, 
we also plot forecasts by Omniscope, which is a future square-kilometer collecting
area array optimized for 21 cm line survey
(Tegmark and Zaldarriaga 2009, Tegmark and Zaldarriaga 2010). \citep{Kohri:2013mxa}
}
\label{fig:CMB+21cm_running}
\end{figure}

In a work studied by K.~Kohri, Y.~Oyama, T.~Sekiguchi, T.~Takahashi,
who are members of SKA-JP (Kohri et al., 2013),
they found that
by using combinations of CMB and 21 cm line we can precisely 
measure the primordial curvature perturbations, which are generated by inflation.
In the previous section,
we take account of higher correlation functions,
but in this work they focus on the two-point correlation function (i.e. power spectrum) 
of the primordial curvature perturbations.
The two point and higher correlation functions 
are complementary to each other,
and both correlation functions are important in order to discriminate inflation models.
In most of previous works about discriminating the inflation models,
mainly focused parameters were
spectral index $n_{s}$ of the power spectrum 
and the dependence of $n_{s}$ on the wave number (i.e. spectral running $\alpha_{s}$).
However, these parameters are not insufficient to distinguish the inflation models in some cases,
and we need higher order wave number dependence of the power spectrum,
e.g. the wave number dependence of $\alpha_{s}$ (i.e. running of running $\beta_{s}$).
Though generally the running of running is very small 
and it is difficult to measure its value,
in that work, they found that
we can strongly improve its constraints 
by combining CMB experiments with precise 21 cm line surveys such as SKA
(see Fig.~\ref{fig:CMB+21cm_running},
and assumed specifications of 21 cm line surveys are listed in Table \ref{tab:21obs_1}).

Also, we can study elementary particle physics by the 21 cm line surveys.
For example, K. Kohri and Y. Oyama
have investigated how well we can constrain properties of neutrino;
the sum of the neutrino masses $\Sigma m_{\nu}$,
the effective number of neutrino species $N_{\nu}$, 
and the hierarchy of neutrino masses (Oyama et al. 2013, Oyama et al. 2015).
Neutrinos are elementary particles with neutral electric charge,
and interact only through the weak interaction with other particles.

Since the discoveries of neutrino masses by Super-Kamiokande 
through neutrino oscillation experiments in 1998, 
the standard model of particle physics has been forced to change to 
theoretically include the neutrino masses. 
So far, only mass-squared differences of neutrinos have been measured 
by neutrino oscillation experiments.
However, the absolute values of neutrino masses and hierarchical structure of them (normal or inverted) 
have not been measured yet
although their information is indispensable to build new particle physics models.
On the other hand, such nonzero neutrino masses affect cosmology significantly 
because relativistic neutrinos erase the gravitational potential up to the free streaming scale. 
Additionally,
the effective number of neutrino species contributes to an extra radiation component
and affects the expansion rate of the Universe.
Besides, the hierarchical structure of neutrino masses
affects both the free-streaming scales and the expansion rate.
Therefore, we can obtain constraints of the neutrino masses, 
the effective number of neutrino species 
and the hierarchy of neutrino masses from cosmological observations.

\begin{figure}[t]
  \begin{center}
   \includegraphics[width=5.5cm]{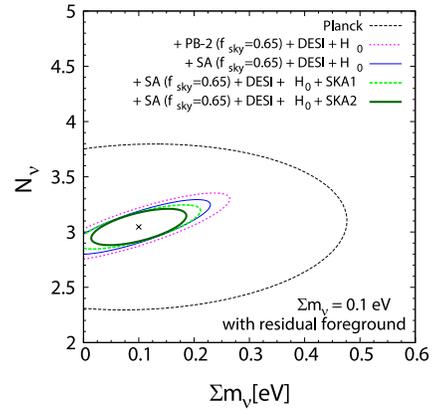} 
  \end{center}
\caption{
Contours of 95\% C.L. forecasts in the $\Sigma m_{\nu}$-$N_{\nu}$ plane
by combinations of CMB (Planck, \textsc{Polarbear}-2 (PB-2), Simons Array (SA))
and SKA (Phase~1 and Phase~2).
Fiducial values of neutrino parameters, $N_{\nu}$ and $\Sigma m_{\nu}$, 
are taken to be $N_{\nu} = 3.046$ and $\Sigma m_{\nu} = 0.1$~eV.
We set the observed redshift range to be $6.75<z<10.05$.
\textsc{Polarbear}-2 and Simons Array are ground-based 
precise CMB polarization observations, 
and DESI is a future baryon acoustic oscillation (BAO) observation. \citep{Oyama:2015gma} 
}
\label{fig:CMB+21cm_Nnu_mass}
\end{figure}

\begin{figure}[t]
  \begin{center}
   \includegraphics[width=6cm]{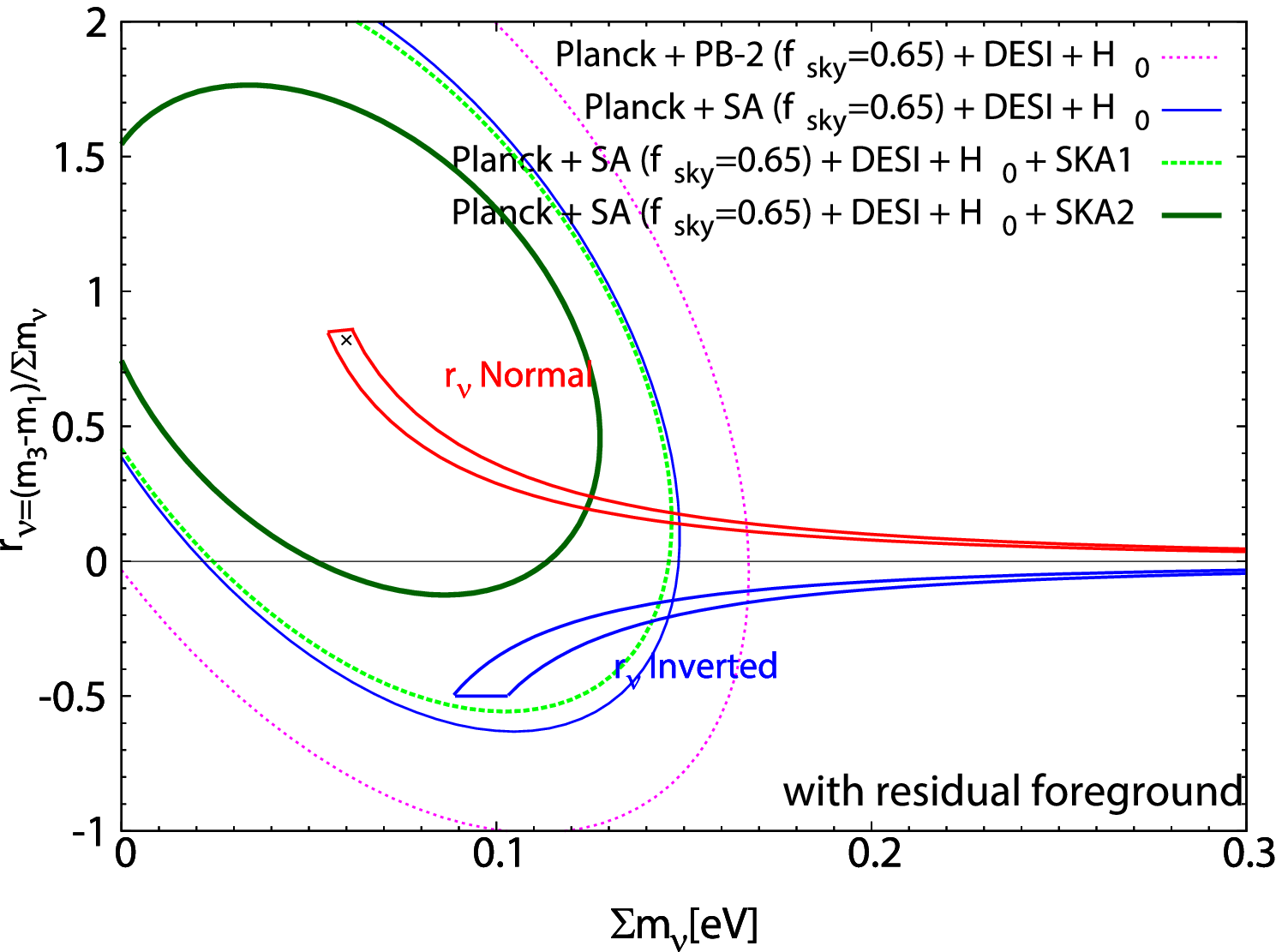} 
   \includegraphics[width=6cm]{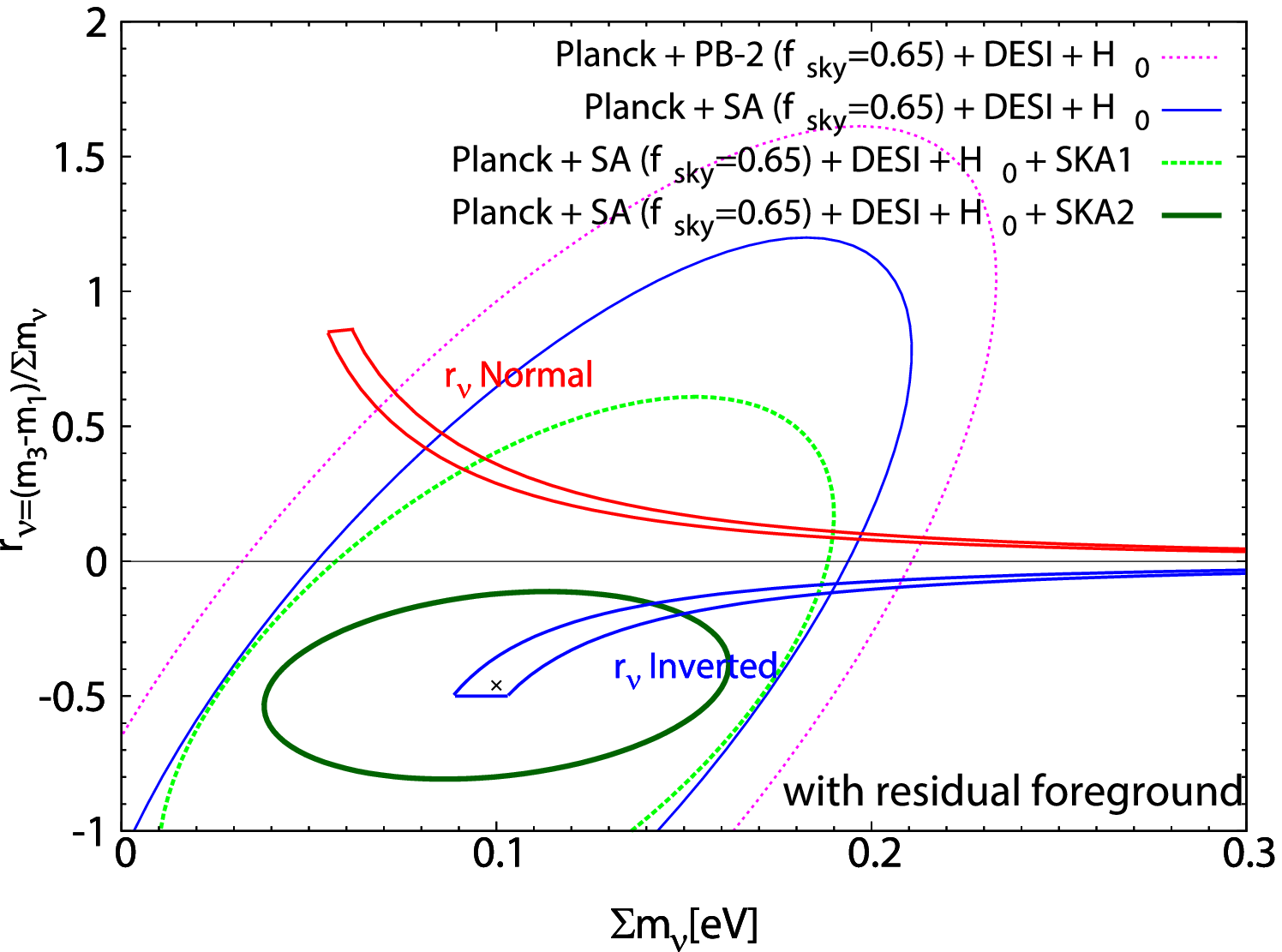}   
\end{center}
\caption{
Contours of 95\% C.L. forecasts in the $\Sigma m_{\nu}-r_{\nu}$ plane
by combinations of CMB (Planck, \textsc{Polarbear}-2 (PB-2), Simons Array (SA))
and SKA (Phase~1 and Phase~2),
where $r_{\nu}\equiv (m_{3}-m_{1})/\Sigma m_{\nu}$.
We assume that an observed redshift range is  $6.75<z<10.05$.
Allowed parameters on $r_{\nu}$ by neutrino oscillation experiments 
are also plotted as two bands for the inverted and the
normal hierarchies, respectively 
(the name of each hierarchy is written in the close vicinity of each band). \citep{Oyama:2015gma} 
}
\label{fig:CMB+21cm_neutrino_hie}
\end{figure}

\begin{figure}[t]
  \begin{center}
   \includegraphics[width=8cm]{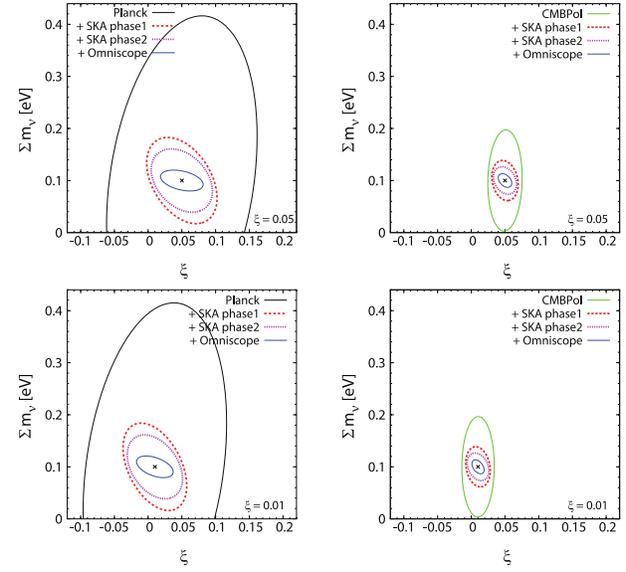} 
  \end{center}
\caption{
Contours of 95\% C.L. forecasts in the $\Sigma m_\nu$--$\xi$ plane,
where $\xi \equiv \mu_{\nu}/T_{\nu}$ represents the lepton asymmetry of the Universe,
$\mu_{\nu}$ is the neutrino chemical potential, and $T_{\nu}$ is the temperature of relic neutrinos.
As CMB data, the Planck and CMBPol surveys are adopted in the left and right panels, respectively.
In order from top to bottom, the fiducial values of $\xi$ are set to be 0.05 and 0.01. 
We assume that an observed redshift range is  $6.75<z<10.05$.
For comparison,  we also plot forecasts by Omniscope. \citep{Kohri:2014hea}
%
}
\label{fig:CMB+21cm_lepton}
\end{figure}

By using SKA1, 
we can strongly constrain on the sum of the neutrino masses $\Sigma m_{\nu}$
and the effective number of neutrino species $N_{\nu}$ (see Fig.~\ref{fig:CMB+21cm_Nnu_mass},
and assumed specifications of 21 cm line surveys are listed in Table \ref{tab:21obs_2}).
In future, by using SKA2,
there is a possibility of distinguishing the hierarchy of neutrino masses
unless the mass structure is degenerated
(see Fig.~\ref{fig:CMB+21cm_neutrino_hie},
and assumed specifications of 21 cm line surveys are listed in the Table \ref{tab:21obs_2}).
Moreover, they found that
we can also obtain strong constraints of the lepton number asymmetry 
of the Universe (Kohri et al., 2014),
which may become a relatively large value in some models of elementary particle physics.
From Fig.~\ref{fig:CMB+21cm_lepton}, 
adding 21 cm line surveys by SKA to CMB observations,
we see that constraints of the lepton number asymmetry
are greatly improved 
in comparison with those of CMB only
(assumed specifications of 21 cm line surveys are listed in Table \ref{tab:21obs_1}).

\begin{figure}[t]
  \begin{center}
   \includegraphics[width=8cm]{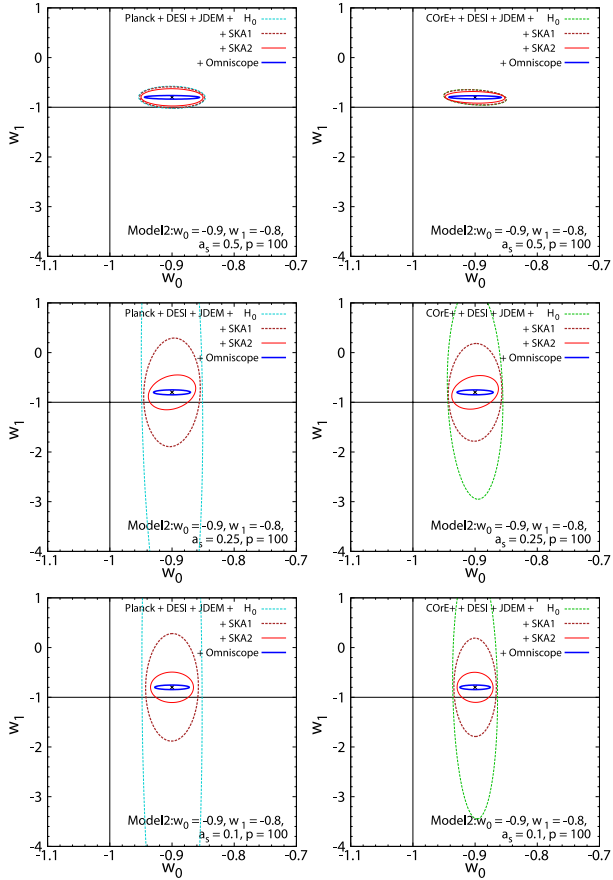} 
  \end{center}
\caption{
Expected constraints at $95\%$ C.L. on the $w_0$--$w_1$ plane for the dark energy EoS parameters.
This EoS is allowed to change at some redshift from $w_1$ (the EoS at earlier time)
to $w_0$ (the EoS at later time), with the extra two parameters to describe
the width $p$ and scale factor $a_s$ of the transition.
The fiducial values are taken to be $w_0=-0.9$, $w_1=-0.8$, $p=100$ and $a_s = 0.5$ (top) , $0.25$ (middle) , $0.1$ (bottom).
Here $a_s$ and $p$ are marginalized over.
In the left and right panels, Planck and COrE+ are assumed as CMB data, respectively. \citep{Kohri:2016bqx} 
}
\label{fig:CMB+21cm_DEEoS}
\end{figure}

Observations of  21 cm line fluctuations can probe different redshift epochs
compared with other cosmological measurements such as CMB, BAO, supernovae, and so on.
Therefore the 21 cm line survey by SKA can also provide us unique information to investigate the nature of dark energy.
Since the dark energy EoS parameter depends on time in most models of dark energy (see Sec.~\ref{sec:DE}), it would be worth
investigating how precisely we can constrain the EoS parameter and its time-dependence.
Considering several parametrizations for the EoS parameter to take the time-dependence into account,
particularly for some parametrization in which the EoS can change at high redshifts,
the future 21 cm observations are found to significantly improve the constraints on the EoS parameters~\citep{Kohri:2016bqx}, as seen in Fig.~\ref{fig:CMB+21cm_DEEoS}.

In these analyses, they set a maximum cut off of spatial scales of
density fluctuations in order to avoid foreground contaminations. Furthermore, in Oyama et al. 2015,
they assume that foreground subtraction can be done down to a given level, and treat contributions of
residual foregrounds as Gaussian random fields.

Thus far, physicists have investigated 
properties of elementary particles by ground-based experiments.
However, as stated above, by using 21 cm line surveys,
we can measure detailed behaviors of matter fluctuations,
and information of them 
enables us to clarify properties of elementary particles. 
This point is specialties of the works stated above.
Therefore, in our research plans,
we aspire for obtaining strong constraints of cosmological parameters
by observing the high redshift Universe 
with the use of the 21 cm line surveys by SKA.
In particular, the members of SKA-JP are world leading scientists 
in the research about particle cosmology by using 21cm line surveys.

In the previous section, 
we focus attention on biased density fluctuations at super horizon scales.
On the one hand, in this section,
we take particular note of the behavior of density fluctuations at relatively small scales,
and both research are complementary to each other.
Also, for the 21 cm line surveys,  we plan to investigate 
how precisely 21 cm line observations can constrain models of the dark energy
and the primordial non-Gaussianity, which have been the focus of the study in the previous section.
Moreover, we also pay attention to
surveys of HI intensity mapping. 
Thus far, research about the HI intensity mapping 
with the motivation  to study elementary particle physics
are not considered so much,
and we expect that this method by using the intensity mapping will develop hereafter.

Also, signals of 21 cm line strongly depend on 
astrophysical processes in the epoch of reionization.
However, we do not know sufficiently
about the reionization history yet.
In addition, 
foregrounds of 21 cm line surveys are very strong radiation
and removal of them is one of the most important subjects.
Members of the reionization group of SKA-JP
particularly focus on 
the astrophysical processes in the epoch of reionization
and the removal of foregrounds.
Therefore, by working with the members of the reionization group,
we can obtain more accurate theoretical models of 21 cm line signals
and methods for the removal of foregrounds.
The models and methods enables us to reduce theoretical uncertainties
in the 21 cm line surveys.

\begin{table}[t]
\centering 
\begin{tabular}{ccccc}
\hline \hline
   & $N_{\rm ant}$& $A_e (z=8)$& $L_{\rm min}$
   & $L_{\rm max}$ \\ 
   & &$[{\rm m}^2]$&$[{\rm m}]$
   & $[{\rm km}]$ \\ 
\hline
SKA1 & $911$  & $443$ & $35$ & $6$  \\ 
Omniscope   & $10^6$ & $1$   & $1$    & $1$ \\ 
\hline \hline 
   & ${\rm FoV}(z=8)$ & $t_{0}$&  $z$ \\ 
   & $[{\rm deg}^2]$&[hour per field]&\\
\hline
SKA1 & $13.12 \times 4 \times 4$  & 4000 & $6.8-10$ \\
Omniscope    & $2.063\times 10^4$ & 16000 & $6.8-10$  \\ 
\hline \hline
\end{tabular}
\vspace{0.1in}
\caption{
Specifications for interferometers of 21 cm experiments adopted in the analyses
(Kohri et al., 2013, and Kohri et al., 2014).
Here, $N_{{\rm ant}}$ is the number of antennae,
$A_e$  is the effective collecting area per a antenna,
$L_{{\rm min}}$ and $L_{{\rm max}}$ are the minimum and maximum baseline, respectively,
${\rm FoV}$ is the field of view, and $t_0$ is the  observation time.
The re-baselining of SKA1 is not considered.
%
For Omniscope, we assume the effective collecting area $A_{e}$ and  field of view are fixed
in the surveyed redshift range.
For SKA2, we assume that
it has the 10 times larger total collecting area than that of SKA1.
%
In addition, for SKA, we assume it uses 4 multi-beaming, 
and its total observation time is the same value as that of Omniscope (16000 hours),
but it observes 4 places in the sky (i.e. 4 times larger FoV and one fourth $t_{0}$).
}
\label{tab:21obs_1}
\end{table}

\begin{table}[t]
\centering 
\begin{tabular}{ccccc}
\hline \hline
   & $N_{\rm ant}$& $A_e (z=8)$& $L_{\rm min}$
   & $L_{\rm max}$ \\
   & &$[{\rm m}^2]$&$[{\rm m}]$
   & $[{\rm km}]$ \\ 
\hline
SKA1  & $911 \times 1/2 $ & $443$ & $35$ & $6$ \\ 
\hline
SKA2  & $911 \times 4   $ & $443$ & $35$ & $6$ \\ 
\hline \hline 
   & ${\rm FoV}(z=8)$ & $t_{0}$&  $z$ \\ 
   &$[{\rm deg}^2]$&[hour per field]&\\
\hline
SKA1  
& $13.12 $  & 1000 & $6.8-10$ \\
\hline
SKA2  
& $13.12 $  & 1000 & $6.8-10$ \\
\hline \hline
\end{tabular}
\vspace{0.1in}
\caption{
%
Specifications for interferometers of 21 cm experiments adopted in the analyses
(Oyama et al., 2015).
We assume that for SKA1 (re-baseline design),
the number of antennae is half as many as
that of the originally planned SKA1, which has 911 antennae,
and for SKA2, 
the number of antennae is 4 times as many as that of originally planned SKA1.
%
Additionally, for SKA 1 and 2, we assume that 
multiple fields are observed by using
these experiments,
and the number of fields is $N_{{\rm field}}=4$ 
in these analyses.
%
%
Then, the effective field of view is
${\rm FoV_{SKA}} = 13.21 \times N_{{\rm field}}$ ${\rm deg}^2$.
}
\label{tab:21obs_2}
\end{table}

\subsection{Cosmic shear measurements with SKA}

Precision weak lensing measurement in cosmology is a key observable for revealing the late-time
evolution of density perturbations.
The weak lensing measurement conducted by the SKA radio continuum survey will probe 
the galaxy population to higher redshift that the lensing signal becomes larger.
The shape (or shear) of galaxies modified by gravitational lensing can be characterized by 
even and odd-parity modes (E- and B-mode, respectively) \citep{Stebbins:1996wx,Kamionkowski:1997mp}, 
defined as
\begin{eqnarray}
	{}_{\pm 2}g({\bf n})=\sum_{\ell m}\left( E_{\ell m}\pm iB_{\ell m}\right){}_{\pm 2}Y_{\ell m}({\bf n})
	\,,
\end{eqnarray}
where ${}_{\pm 2}g$ denotes the spin$-\pm 2$ component of the reduced shear field, 
which can be expanded by the spin$-\pm 2$ spherical harmonics ${}_{\pm 2}Y_{\ell m}$.
Here $E_{\ell m}$ and $B_{\ell m}$ represent the two parity eigenstates with electric-type $(-1)^\ell$
and the magnetic-type $(-1)^{\ell +1}$ parities, respectively.
The symmetric argument implies that the density perturbations at linear-order produce only the E-mode 
of the cosmic shear field (see e.g. \cite{Yamauchi:2013fra}).
Hence, non-vanishing B-mode signals on large angular scales immediately imply the presence of
non-scalar perturbations.
Although the amplitude of these signals is small, it is known that there are a variety of generating mechanism. 
Possible subdominant components in the Universe, such as active vector modes generated by topological defects~\citep{Yamauchi:2012bc}, 
primordial gravitational waves~\citep{Dodelson:2003bv}
and secondary vector/tensor modes generated by density perturbations~\citep{Saga:2015apa,Sarkar:2008ii}, rather than
first-order density perturbations, can induce the B-mode shear.
A search for those tiny signals from the B-mode shear is a key to explore subdominant components of the Universe.

\begin{figure}
	\begin{center}
		\includegraphics[width=8cm]{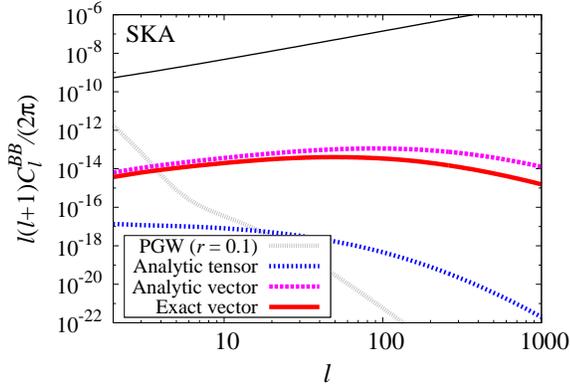} 
	\end{center}
		\caption{
			Angular power spectrum of the weak lensing B-mode.
			For comparison, the black solid line shows the expected statistical error estimated by the shot noise
			for the weak lensing measurement conducted by SKA2.
			The second-order vector mode dominates the expected signals on small scales~\citep{Saga:2015apa}.
		}\label{Fig:cls_BB_SKA}
\end{figure}

The angular power spectra of the B-mode cosmic shear, which is defined through 
$C_\ell^{\rm BB}=\sum_m\langle |B_{\ell m}|^2\rangle /(2\ell +1)$, 
generated by primordial gravitational waves and the second-order vector/tensor modes induced by the linear-order density perturbations 
are shown in Fig.~\ref{Fig:cls_BB_SKA}, implying that the total B-mode shear signals cannot
exceed the expected noise level even for the weak lensing measurement conducted by the SKA2~\citep{Saga:2015apa},
while the signal induced by the second-order vector mode turns out to dominate on almost all scales.
On the other hand, the weak lensing measurement in the HI IM survey is expected to substantially
improve the signal-to-noise ratio and provide the observable signals, because 
the angular power spectrum of the brightness temperatures
do not have diffusion scale unlike CMB and measures the fluctuations at different epochs.
We leave this for future investigations.

\subsection{Synergy between SKA and CMB experiments: primordial gravitational-waves}

Measurements of the odd-parity pattern (B-modes) in the CMB polarization
on large angular scales have been considered as the best avenue to probe the primordial 
gravitational-waves as a smoking gun of the cosmic inflation \citep{Kamionkowski:1996zd}. 
Still, there is no evidence for the presence of the primordial gravitational-waves, and
the amplitude of the primordial gravitational-waves relative to the primordial density fluctuations, 
usually referred to as the tensor-to-scalar ratio $r$, is constrained as 
$r<0.07$ ($95$\% C.L.) from the Joint analysis of {\sc BICEP2}, {\it Keck Array} and 
{\it Planck} Collaborations \citep{Array:2015xqh,Ade:2015tva}. 
The detection of the gravitational-wave (primary) B-modes is one of the main targets in ongoing and 
future CMB experiments.

On large angular scales, B-modes from Galactic foreground emission are expected to dominate over 
the primary B-modes. Multiple studies (e.g., \cite{Dunkley:2008am,Betoule:2009pq,Katayama:2011eh}) 
have been devoted for foreground cleaning techniques (see \cite{Ichiki:2014qga} for a resent review).
\citet{Katayama:2011eh} showed that the foreground contamination would be possible to be subtracted
enough to detect the primordial B-modes at $r\sim 0.001$.
However, even if the foreground contamination is successfully mitigated, 
there is still a significant contamination from the gravitational lensing effect on
the CMB polarization caused by the gravitational potential of 
the large scale structure \citep{Zaldarriaga:1998ar}. 
Precise estimate and subtraction of the lensing B-modes, usually referred to as {\it delensing}
\citep{Seljak:2003pn,Smith:2010gu,Namikawa:2015tba}, will be required in ongoing and 
future CMB experiments such as \textsc{BICEP}/{\it Keck Array} series \citep{Ade:2015fwj,Array:2016afx} 
and LiteBIRD \citep{Matsumura:2013aja}. 

The lensing B-modes can be estimated from measurements of the CMB-lensing mass fields or 
using Cosmic Infrared Background (CIB) intensity map. The CMB-lensing mass fields have been 
precisely reconstructed from CMB polarization experiments 
\citep{Ade:2013gez,vanEngelen:2014zlh,Story:2014hni,Ade:2015zua} (see \cite{Namikawa:2014xga} for a recent review), 
and the lensing B-modes estimated from the CMB-lensing mass fields 
have been detected with high significance very recently \citep{Ade:2015zua}.
Detections of the lensing B-modes from the CIB observations have been also reported 
in several CMB experiments \citep{Hanson:2013hsb,vanEngelen:2014zlh,Ade:2015zua}.
These lensing B-modes are, however, measured at scales where the primary B-modes have negligible contribution.
The delensing with these lensing B-modes therefore does not improve the sensitivity to the primary B-modes. 

\begin{figure}
\begin{center}
\includegraphics[width=8cm]{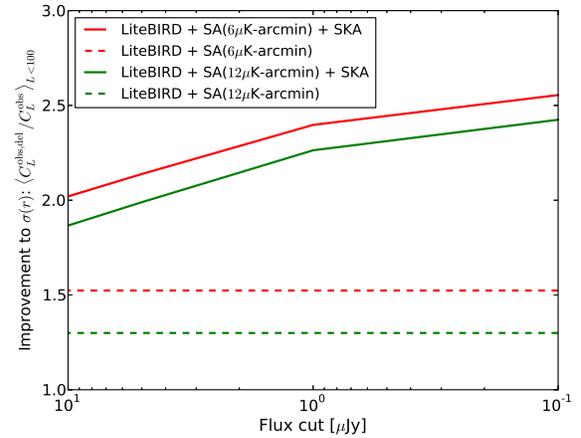} 
\end{center}
\caption{
Delensing improvement to constraints on the tensor-to-scalar ratio as a function of
the flux cut of the SKA radio continuum survey. 
The solid lines show the delensing improvement by combining the LiteBIRD, Simons Array (SA) and SKA, 
while the horizontal dashed lines show the case with CMB experiments alone.
The red (green) lines show the case if the polarization sensitivity of the Simons Array 
becomes $6\mu$K-arcmin ($12\mu$K-arcmin). (see \cite{Namikawa:2015tjy})
}
\label{Fig:delensing}
\end{figure}

The estimate of the lensing B-modes would be also possible using any CMB-lensing mass tracers
such as halos/galaxies at high redshifts which generate the most part of the gravitational potential of the CMB lensing 
\citep{Marian:2007sr,Simard:2014aqa,Sherwin:2015baa}.
The SKA radio continuum survey \citep{Jarvis:2015tqa} will provide a two-dimensional intensity map 
of the synchrotron radiations from halos/galaxies integrated along the line of sight. 
These sources would be located at high redshifts ($z\sim 2$).
Since the CMB lensing is mostly generated by inhomogeneities of 
the mass distribution at such redshifts, fluctuations in the intensity map measured from the radio continuum survey 
would effectively trace the gravitational potential of the CMB lensing. 
The intensity map is therefore possible to be used for
estimating part of the lensing B-mode contributions, 
and subtracting it from the observed B-modes on large scales. 

\citet{Namikawa:2015tjy} explored expected delensing performance in future CMB experiments 
using the intensity map of the SKA radio continuum survey.
Fig.~\ref{Fig:delensing} shows the improvement to the constraints on the tensor-to-scalar ratio 
of the LiteBIRD by performing the delensing analysis (for details, see \cite{Namikawa:2015tjy}).
We assume that the large-scale lensing B-modes of the LiteBIRD are {\it delensed} using
the CMB-lensing mass fields reconstructed from the LiteBIRD and Simons Array (SA) \citep{2014SPIE.9153E..1FA}, 
and the intensity map of the SKA radio continuum survey. 
The improvement is computed with varying the flux cut of the radio continuum survey. 
Since the future polarization sensitivity of the SA is uncertain, 
we show two possible cases of the SA polarization sensitivity.
The inclusion of the SKA1-delensing will significantly increase the improvement to the constraints on 
the tensor-to-scalar ratio by $80$--$120\,\%$ compared to the case without the delensing analysis. 
With the SKA2 ($0.1$--$1\mu$Jy) and CMB observations, 
the improvement to the constraints becomes $130$--$160$\,\%. 
Note that the additional improvement by the inclusion of the SKA2 to the CMB data is approximately $100$\,\%. 
The intensity map from the SKA radio continuum survey would be quite useful for the future CMB delensing analysis, 
especially for the LiteBIRD observation.

\section{Summary}

In summary, the SKA will yield transformational science across a wide range of cosmology in the next decades.
In particular, the SKA has several cosmological surveys such as
the HI galaxy, HI intensity mapping and radio continuum survey, which have the potential
to open the new frontier of cosmology and deliver the precision cosmology (see Sec.~\ref{sec:surveys}).
The cosmology science that the SKA has the potential to provide is of great interest 
for SKA-Japan consortium (SKA-JP) cosmology science working group.
In this paper, we have briefly reviewed the cosmology science with the SKA and 
have highlighted some examples of specific topics that Japanese cosmologists
are currently playing an important role.

Our analysis has revealed that the SKA should possess the extremely-large survey volume, e.g. all-sky,
to enhance the study of the mechanism for generating the primordial density fluctuations 
through the scale-dependent bias (see Sec.~\ref{sec:multitracer technique}).
In this treatment, the precise estimate of halo mass of each galaxy inferred from available 
observables is quite essential to take the advantage of the multitracer technique.
Hence, the properties of individual galaxies, such as mass and redshift, 
inferred from available data by other surveys may provide the further improvement in the constraints.

We also expect that the SKA should explore the very high-redshift dark Universe,
such as the cosmic dawn and dark age, with the $21$-cm surveys (see Sec.~\ref{sec:dark Universe}).
However, as already mentioned, the signals of the $21$-cm line during the EoR strongly depends 
on the astrophysical processes, which are poorly understood and include the large uncertainties.
We should further improve the cleaning method to remove the foregrounds by working with the members
of the reionization group.

Finally, we emphasize that in the SKA era the cosmological surveys will be limited by systematic
errors and cosmic-variance noises rather than statistical errors because future planned surveys
will be able to probe the huge number of samples.
This implies that the synergy with other wavelengths is quite important
and possible synergy with multiwavelength data from e.g., optical/infrared and CMB measurements,
will have the potential to reduce these noises.
The Japanese astronomical researchers are currently leading cosmology-oriented surveys 
in other wavelengths such as HSC and LiteBIRD, and participating in ongoing and upcoming experiments.
By working with members of these surveys, the Japanese community has the potential to 
provide the powerful tools for cosmology in next decade.
The involvement of SKA-JP in the SKA will benefit our community by enlarging their expertise in cosmology science.

\section*{Acknowledgements}

The authors are grateful to International SKA cosmology science working group members for providing us 
opportunities of open discussion and cooperation.
We would like to thank M.~Oguri for useful discussions.
We also thank T.~Akahori and T.~Takeuchi for carefully reading our manuscript.
D.Y. is supported by Grant-in-Aid for the Japan Society for the Promotion of Science (JSPS) Fellows (No.~259800). 
T.N. is supported by JSPS fellowshop for abroad (No.~26-142).
The works of K.K., K.T., and T.T. are partially supported by Grand-in-Aid from the JSPS and 
the Ministry of Education, Culture, Sports, Science and Technology (MEXT) of Japan, 
No.~26105520 (K.K.), No.~26247042 (K.K.), No.~15H05889 (K.K.),
No.~24340048 (K.T.),  No.~26610048 (K.T.), No.~15H05896 (K.T.), 
No.~15K05084 (T.T.), No.~15H05888 (T.T.).
The work of K.K. is also supported by the Center for the Promotion of Integrated Science
(CPIS) of Sokendai (1HB5804100).


\end{document}